# A Structured Framework for Assessing the "Goodness" of Agile Methods

Shvetha Soundararajan and James D. Arthur

Department of Computer Science
Virginia Tech
Blacksburg, VA 24061
{shvetha, arthur}@vt.edu

*Abstract – Agile Methods are designed for customization; they offer an organization or a team the flexibility to adopt a set of principles and practices based on their culture and values. While that flexibility is consistent with the agile philosophy, it can lead to the adoption of principles and practices that can be sub-optimal relative to the desired objectives. We question then, how can one determine if adopted practices are "in sync" with the identified principles, and to what extent those principles support organizational objectives? In this research, we focus on assessing the "goodness" of an agile method adopted by an organization based on (1) its adequacy, (2) the capability of the organization to provide the supporting environment to competently implement the method, and (3) its effectiveness. To guide our assessment, we propose the Objectives, Principles and Practices (OPP) framework. The design of the OPP framework revolves around the identification of the agile objectives, principles that support the achievement of those objectives, and practices that reflect the "spirit" of those principles. Well-defined linkages between the objectives and principles, and between the principles and practices are also established to support the assessment process. We traverse these linkages in a top-down fashion to assess adequacy and a bottom-up fashion to assess capability and effectiveness. This is a work-in-progress paper, outlining our proposed research, preliminary results and future directions.*

*Key words: Agile Assessment, Adequacy, Capability, Effectiveness, Objectives, Principles, and Practices*

## I. INTRODUCTION

The agile philosophy provides an organization or a team with the flexibility to adopt a selected subset of principles and practices based on their organizational culture, values and the type of system under consideration. More specifically, organizations or teams follow customized agile methods, tailored to better accommodate their needs. However, the extent to which a customized approach satisfies the needs of an organization, or rather the "goodness" of that approach, should be confirmed. Existing methods to assess the "goodness" of agile methods focus on a comparative analysis or are limited in scope and application. For example, the burn-up and burn-down charts are used by agile teams to indicate the amount of work done or the amount of work remaining. They do not, however, indicate if the development approach is effective or if their people are skilled workers. Most assessment approaches for agile methods focus on assessing the working software and process artifacts. In particular, they place emphasis on the *product*. Nonetheless, that is not to say that there are no approaches to assess the *process*. There exist Agile Process Improvement Frameworks such as the Sidky Agile Measurement Index (SAMI) [1, 2] and Agile Adoption and Improvement Model (AAIM) [3] that guide an organization's agile adoption and improvement efforts. Both frameworks describe levels of agility modeled on similar concepts found in the SW CMM [4] and CMMI [5]. That is, a set of practices is to be adopted by an organization at each level in order to be "agile" at that level. The primary disadvantage of these frameworks is that a set of practices is "forced" on an organization at defined levels, which compromises the flexibility offered by agile methods.

We advocate the need for a more comprehensive agile assessment process that assesses the *people, process, project and product* characteristics of organizations adopting agile methods. In this research, we propose an approach to assess the "goodness" of agile methods from three perspectives. More specifically, to assess the "goodness" of a given agile method, we seek to address the following three questions:

- How *adequate* is the method with respect to achieving its objectives?
- How *capable* is an organization in providing the support environment to implement its selected method?
- How *effective* has been the implementation of the method in achieving its objectives?

In response to the above questions, we have developed the **O**bjectives, **P**rinciples and **P**ractices (OPP) framework to facilitate the assessment of the adequacy, capability and effectiveness of agile methods. The framework identifies desirable objectives embraced by the agile philosophy, and definitively links them to principles that support the achievement of those objectives. Similarly, accepted practices are identified and linked to the principles that they support.

The linkages between the objectives and principles, and between the principles and practices, guide the assessment process. We assess the *adequacy* of an agile method by traversing the linkages in a top-down fashion. That is, given the set of objectives espoused by the agile method, we follow the linkages downward to ensure that the appropriate principles are enunciated, and that the proper practices are expressed. In addition to a top-down examination, the

*capability* of an organization to implement its adopted methodology and its *effectiveness* are assessed using a complementary bottom-up traversal of the linkages. This begins, however, by identifying people, process, project and product characteristics (or *indicators*) that attest to the use of particular practices. Then, by following the linkages upward from the practices, we can infer the use of proper principles and the achievement of desired objectives.

Section 2 describes our proposed approach. It provides an overview of the OPP framework and its components. Our approach to assessing "goodness" is discussed in Section 3. We present our proposed substantiation approach in Section 4. In Section 5, we outline what we have accomplished so far. Section 6 summarizes our work.

## II. EVOLVING THE ASSESSMENT FRAMEWORK

Our research is motivated by the lack of a comprehensive approach to assessing agile methods. We assess the "goodness" of an agile method adopted by an organization based on (1) its *adequacy*, (2) the *capability* of the organization to provide the supporting environment to implement the method, and (3) its *effectiveness*. We define adequacy, capability and effectiveness as below:

- Adequacy - Sufficiency of the method with respect to meeting stated objectives
- Capability – Ability of the organization to provide the supporting environment conducive to the implementation of the method - which is dependent on its people, process and project characteristics.
- Effectiveness – Producing the intended or expected results - which is dependent on process and product characteristics

### A. The Framework

Figure 1 provides an overview of the OPP framework. The **OPP** framework identifies (1) objectives of the agile philosophy, (2) principles that support the objectives, (3) practices that are reflective of the principles, (4) the linkages between the objectives, principles and practices and (5) indicators to assess the characteristics of the people, process, project and product.

The culture of an organization, its values and desired characteristics of the systems that it builds determine the objectives, principles and practices that it adopts. Our assessment of an agile method is carried out with respect to satisfying its objectives. Figure 1 illustrates the relationships among the objectives, principles, practices and indicators. This relationship is central to our assessment process and is common to the assessment of adequacy, capability and effectiveness. Within an organization, to assess the adequacy of an agile method, we traverse the linkages in a top-down manner from objectives to principles and from principles to practices. The existence of principles and practices supporting the desired objectives is indicative of the adequacy of the method under consideration.

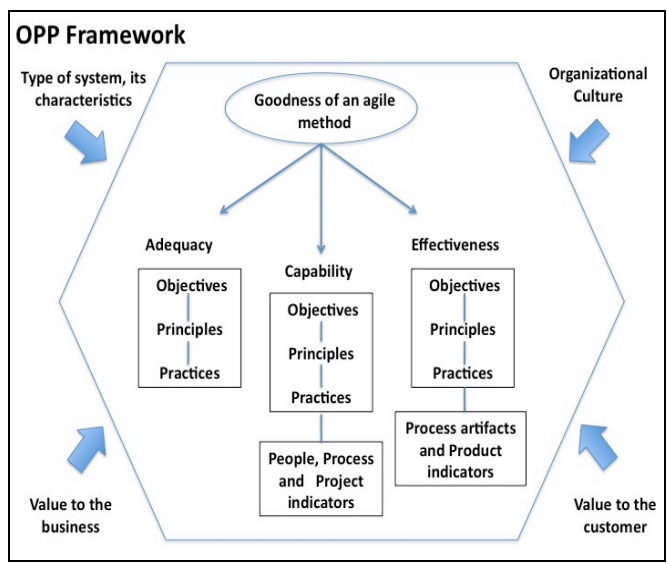

Figure 1. The Objectives, Principles, Practices (OPP) Framework

Additionally, the OPP framework identifies people, process, project and product characteristics of the practices adopted. These observable characteristics that are associated with the practices are called indicators. These indicators are essential to the assessment of capability and effectiveness. Figure 1 also shows the linkages between practices and indicators for assessing capability and effectiveness. *People*, *process* and *project* indicators imply the presence or absence of characteristics needed in the supporting environment, i.e., capability. Similarly, additional *process artifacts* and *product* indicators denote expected results. They are used to effect a bottom-up assessment of effectiveness by traversing the linkages from the appropriate indicators to practices, practices to principles, and principles to objectives.

### B. Formulated Components of the OPP Framework

At the heart of the OPP framework are the objectives, principles, practices and the linkages that tie them together. Indicators are identified and are required for the assessment of capability and effectiveness. The tasks necessary to sufficiently define the components of OPP framework are as follows:

1. *Deriving the objectives and identifying the supporting principles and the practices*

The agile manifesto [6] provides four focal values and twelve principles that define the agile philosophy. Our work involves deriving objectives reflective of the agile philosophy from the focal values, identifying principles that support the defined objectives, and identifying practices that reflect the principles.

As illustrated in Figure 2, we have identified an initial set of objectives based on the agile philosophy. A supporting aggregated set of principles have also been identified from sources including, but not limited to, the agile manifesto, books, research papers, experience reports, white papers and discussions with industry experts. Likewise, we have identified a list of practices embraced by the agile community.

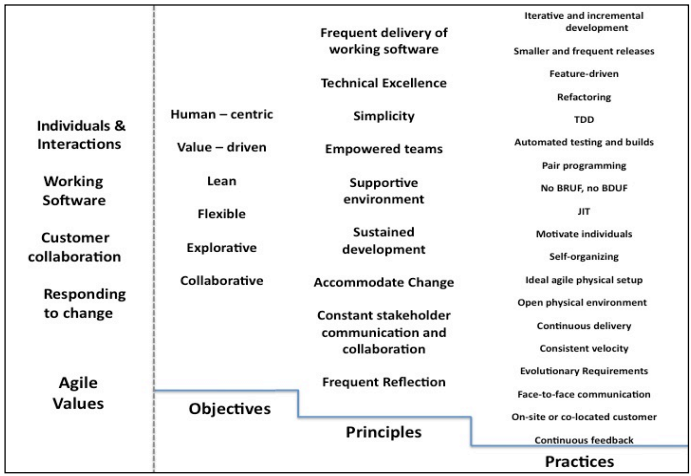

Figure 2. Objectives, Principles and Practices

These objectives, principles and practices are used to the assessment of adequacy, capability and effectiveness.

2. *Establishing definitive linkages between the identified objectives, principles and practices*

Linkages between the objectives, principles and practices have to be defined in order to assess the adequacy, capability and effectiveness. We propose to gather evidence for the existence of the defined linkages from the literature, experience reports, and analysis of existing agile methods. Example linkages are shown in Figure 3 below.

Let us assume that an organization lists *flexibility* as one of its objectives. One underlying principle that supports flexibility is *accommodating change*. Hence there exists a linkage between the objective "flexible" and the principle "accommodate change." We then have a set of practices such as *face-to-face communication, on-site or co-located customer* and *no BRUF* that help realize the principle of accommodating change. The traversal approaches discussed here reflect a hierarchical structure and can be supported by the Evaluation Environment [7].

Although not shown in figure 3, the OPP framework supports an additional level of linkages between practices and indicators that are established to assess capability and effectiveness. Those indicators are discussed next.

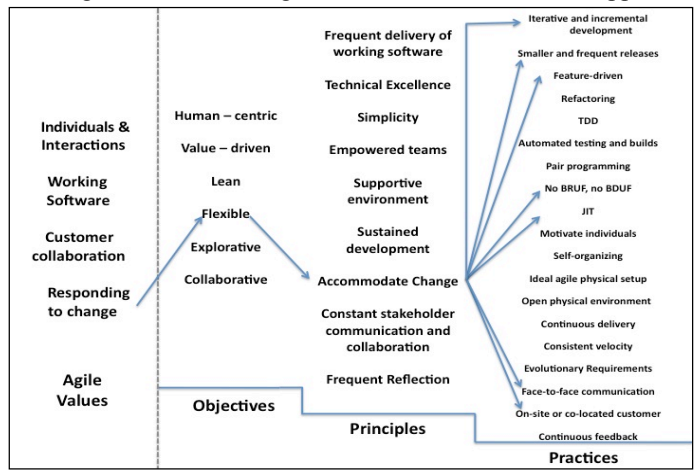

Figure 3. Example Linkages in the OPP Framework

3. *Identifying indicators*

Indicators are observable characteristics of the people, process, project and product, and are linked directly to practices employed by an organization. We propose to identify and substantiate the indicators by reviewing the literature.

Indicators are required to assess

 i. the capability of the organization, and
 ii. the effectiveness of the method

People, process and project indicators that denote the characteristics of the environment are used to assess the capability of the organization. Assessing the effectiveness of the agile method involves the identification of process artifacts and the product indicators, which focus on "working software".

### III. ASSESSING "GOODNESS"

As mentioned earlier, we assess the "goodness" of an agile method by assessing its adequacy, the capability of the organization to provide the supporting environment to implement the method, and its effectiveness.

*A. Assessing Adequacy*

Adequacy is defined by the sufficiency of an agile method with respect to meeting stated objectives. It is *independent* of an organization. More specifically, we can assess the adequacy of standalone agile methods such as XP, Scrum and FDD with respect to the agile values and principles each espouses. That is, given an objective of the method, are the necessary principles also present that are prescribed by the framework? Then, for each principle enunciated by the framework, are practices that are prescribed by the framework present within the agile method? If necessary principles and practices are missing, then adequacy is suspect.

*B. Assessing Capability*

Capability is the ability of the organization to provide the supporting environment to implement its adopted agile method. This depends on the people, process and project characteristics of an organization, and are reflective of the environment within which an agile method is employed.

To assess capability, we propose a bottom-up traversal of the linkages from the indicators to the objectives (see Figure 1). Before assessing the capability of an, however, the adequacy of the method must also be determined.

*C. Assessing Effectiveness*

Effectiveness of a method lies in the ability of an organization to produce the intended or expected results. Assessing the effectiveness also involves a bottom-up traversal from process artifacts and product quality indicators to the objectives.

Similar to assessing capability, we must also assess the adequacy of the adopted agile method.

On a final note, unlike adequacy, both capability and effectiveness are assessed from an organizational perspective. Hence, we cannot assess the capability and effectiveness of a standalone agile method that is independent of an organization.

## IV. SUBSTANTIATING THE ASSESSMENT FRAMEWORK

The OPP framework guides the assessment process. Our goal is to substantiate both the components of the OPP framework and our process for assessing adequacy, capability and effectiveness.

### A. Substantiating the components of the OPP Framework

The objectives, principles, practices, and the linkages between them, form the core of the OPP framework (see Figures 1, 2 and 3). The framework also identifies people, process, project and product characteristics that are necessary to assess the capability of the organization and the effectiveness of the method under consideration. We outline the following 2-step substantiation approach:

a) Validate the components of the OPP framework by obtaining feedback from agile practitioners using survey instruments and interviews.

b) Gathering evidence from literature to
   i. validate the linkages between objectives, principles and practices, and
   ii. confirm the indicators.

We are currently involved in gathering evidence from research papers, experience reports, white papers and books to validate the existence of the proposed linkages.

### B. Substantiating the assessment process

We address the assessment of agile methods from three perspectives – adequacy, capability and effectiveness. We realize that in order to effectively substantiate our assessment approach, the OPP framework has to be applied within an organization.

a) Using the OPP framework, we first propose to assess the adequacy of multiple agile methodologies endorsed by the agile community (XP, Scrum, Lean, etc.).

b) Secondly, we intend to apply the OPP framework within multiple organizations to assess
   i. the adequacy of its agile method
   ii. the capability of the organization to provide the supporting environment to implement that agile method, and
   iii. the effectiveness of its agile method

While necessary, item (iii) requires a longitudinal study that falls beyond the scope of our immediate substantiation goals.

## V. PRELIMINARY RESULTS

We have identified the core set of objectives, principles and practices, and are in the process of establishing the linkages among them. As mentioned earlier, assessment of adequacy can be carried out on standalone agile methods. We are working towards evaluating the adequacy of XP, Scrum and FDD with respect to the OPP Framework. Our results and analysis are not yet complete and reporting them at this stage will be premature.

## VI. SUMMARY

Our research has been motivated by the need for a comprehensive approach to assess the "goodness" of agile methods. We assess "goodness" based on (a) a method's adequacy, (b) the capability of an organization to provide the supporting environment for implementing the method, and (c) the effectiveness of its method. The OPP framework defines objectives, principles, practices and indicators, and linkages between them to support the assessment process. Our proposed substantiation approach includes a study of one or more organizations to assess the "goodness" of their agile methods, at least from an adequacy and capability perspective.